\documentclass[preprint,preprintnumbers,showpacs,aps,prd,amssymb]{revtex4-1}

\usepackage{graphicx}
\usepackage{bm}
\usepackage{amsmath}

\def\calO{{\cal O}}
\def\calN{{\cal N}}

\def\nn{\nonumber}

\begin{document}
\title{Quantization of $D$-dimensional noncommutative black holes}
\author{Jong-Phil Lee}
\email{jplee@kias.re.kr}
\affiliation{Institute of Convergence Fundamental Studies, Seoul National University of Science and Technology, Seoul 139-931, Korea}
\affiliation{Division of Quantum Phases $\&$ Devices, School of Physics, Konkuk University, Seoul 143-701, Korea}

\begin{abstract}
Noncommutative black holes in higher dimensions are investigated in the context of holographic principle.
Quantization rules for the discrete mass spectrum are derived and compared with the continuous spectrum in the literature.
Because of the noncommutative nature of background geometry the minimum mass to form a noncommutative  black hole is very large
(it becomes larger for discrete spectra), 
so the current LHC search results for mini black holes cannot exclude the possibility of quantum gravity in higher dimensions.
\end{abstract}

\maketitle
Successful runnings of the Large Hadron Collider (LHC) at CERN open a new era of high energy physics.
It culminated in a discovery of a new boson consistent with the Standard Model (SM) Higgs particle recently, 
and a TeV scale regime is probed for the first time.
One of the most fascinating events that might be witnessed at the LHC is the production of mini black holes.
Mini black holes can be produced at the LHC if there exist extra dimensions \cite{XD}.
This is because the fundamental energy scale for strong gravitation $M_*$ in $D=(4+n)$-spacetime dimensions
is much smaller than the Planck mass.
For a mass scale of $M_*\sim 1$ TeV, the LHC is expected to produce $D$-dimensional mini black holes \cite{BH}.
Typically the size of the Schwarzschild radius of mini black holes is $\sim\calO(10^{-4})$ fm.
Mini black holes also evaporate through Hawking radiation just as ordinary black holes, 
and the lifetime is $\sim 10^{-26}~\sec$.
Thus once a black hole is produced at the LHC, its creation could be detected by Hawking radiation.
It is also possible that rotating or charged black holes could be produced at the LHC.
Some of the interesting features of rotating black holes are studied in \cite{jplee0}.
For recent searches for mini black holes at the LHC, see \cite{CMS,ATLAS}.
\par
If there exists scale invariance at some high energy scale, its effect is described by unparticles.
Tensor unparticles can produce the so called ungravity.
In a strong ungravity region, black holes induced by ungravity can be formed.
Unparticle black holes look much like higher dimensional black holes,
but with fractional number of dimensions.
One can distinguish ungravity black holes from ordinary ones by investigating their quasi-normal modes \cite{jplee1}.
\par
However, all these black holes are classical in the sense that they show singular behavior at the origin.
For example, the Hawking temperature is inversely proportional to the black hole mass so the temperature becomes
infinitely large at the final stage of evaporation via Hawking radiation.
It is widely believed that some kind of quantum rules would work out for taming the singular behavior of black holes.
One of the most promising candidate for describing quantum nature of spacetime is noncommutative(NC) geometry \cite{Witten, Seiberg}.
Black hole physics in the background of the NC geometry has already been studied comprehensively so far
\cite{Nicolini05,Nicolini06,Myung06, Ansoldi06,Mukherjee,Banerjee08,Banerjee082,Arraut09,Mukherjee2,Banerjee09,Smailagic10, Mureika11}.
Higher dimensional NC black holes are also investigated in \cite{Rizzo06,Nicolini08,Nicolini11}.
\par
For ordinary Schwarzschild geometry the source mass is considered to be distributed as a delta function.
But in a NC geometry the mass density is distributed in a Gaussian form with a dispersion of order $\sim\theta$
which measures the noncommutativity of spacetime \cite{Smailagic, Smailagic2}:
\begin{equation}
 \rho_{cont} (r)=\frac{M}{(4\pi\theta)^{3/2}}\exp (-r^2/4\theta)~,
\label{rhocont}
\end{equation}
where $M$ is the black hole mass parameter.
In a NC geometry there is a fundamental uncertainty in a small region of order $\sim\sqrt{\theta}$, 
below which we cannot probe or specify precise location.
This is the reason why mass density is fuzzy in the form of Eq.\ (\ref{rhocont}), not as a delta function.
One of the most important results of this setup is that there is a minimum or threshold mass 
(and consequently minimum horizon radius).
Below the threshold, the event horizon does not exist.
At the minimum horizon radius the Hawking temperature vanishes and no singular behavior occurs.
\par
In a recent work the author provided with the quantization rule for the NC black holes \cite{jplee2}.
The work was based on the assumption that the NC spacetime geometry inside
every black hole is quantized in such a way that the surface area is quantized in units of the
minimum area given by the minimum black hole radius.
This is a kind of holographic principle where all the relevant information about the black hole is {\em ``pixelated''} on its surface.
In short the black hole surface area is assumed to be quantized as
\begin{equation}
 A_n=4\pi r_h^2=4\pi r_0^2 n~,
 \label{An4}
\end{equation}
where $r_h$ is the horizon radius, $r_0\sim ({\rm a~few})\times\sqrt{\theta}$ is the minimum horizon radius,
and $n=1, 2, \cdots$.
Ref. \cite{Spallucci} adopts a similar quantization.
The quantization rule for $r_h$ is then 
\begin{equation}
 r_h=r_0\sqrt{n}~.
\label{rn4}
\end{equation}
Note that these quantizations are for $D=4$ dimensions.
\par
In this work we generalize the idea of quantization of NC black holes into that of $D$-dimensional ones.
In $D$ dimensions the quantization of Eq.\ (\ref{An4}) becomes
\begin{equation}
A_n=\frac{2\pi^{(D-1)/2}}{\Gamma\left(\frac{D-1}{2}\right)}r_h^{D-2}
=\frac{2\pi^{(D-1)/2}}{\Gamma\left(\frac{D-1}{2}\right)}r_0^{D-2}\cdot n~,
\label{An}
\end{equation} 
and thus
\begin{equation}
r_h=\left(n^\frac{1}{D-2}\right)r_0~.
\label{rn}
\end{equation}
Using the $D$-dimensional quantization of Eqs.\ (\ref{An}) and (\ref{rn}), we derive the quantization rules
for $D$-dimensional NC black holes in what follows.
As for the quantization of ordinary $D$-dimensional mini black holes, see \cite{Dvali}.
\par
In discretized spacetime, the mass density of Eq.\ (\ref{rhocont}) now becomes
\begin{equation}
\rho(r)=\frac{M}{N_0}e^{-r^2/4\theta}~.
\label{rho}
\end{equation}
Here the normalization $N_0$ is
\begin{equation}
N_0=\frac{2\pi^{(D-1)/2}}{\Gamma\left(\frac{D-1}{2}\right)}r_0^{D-1}
\cdot\sum_{n=1}^\infty n\cdot\exp\left[-\alpha\cdot n^{2/(D-2)}\right]~,
\label{N0}
\end{equation}
where $\alpha\equiv r_0^2/(4\theta)$.
In continuous spacetime the metric function is 
\begin{equation}
h_{cont}(r)
=1-\frac{1}{r^{D-3}}\left(\frac{1}{\sqrt{\pi}M_*}\right)^{D-3}
 \frac{8\Gamma\left(\frac{D-1}{2}\right)}{D-2}\frac{m_{cont}(r)}{M_*}~,
 \end{equation}
where
\begin{eqnarray}
m_{cont}(r)
&=&
\int_0^r\rho_{cont}(r)
\frac{2\pi^{\frac{D-1}{2}}}{\Gamma\left(\frac{D-1}{2}\right)}~r^{D-2}dr\nn\\
&=&
\frac{M}{\Gamma\left(\frac{D-1}{2}\right)}
\gamma\left(\frac{D-1}{2},\frac{r^2}{4\theta}\right)~.
\end{eqnarray}
But for discrete spacetime, one should follow the quantization rule of Eq.\ (\ref{rn})
and $m_{cont}(r)$ must be replaced by
\begin{equation}
m(N)=\sum_{n=1}^N\rho(r)
\left[\frac{2\pi^{\frac{D-1}{2}}}{\Gamma\left(\frac{D-1}{2}\right)}\right]
r^{D-2}\cdot\Delta r~,
\label{mN1}
\end{equation}
where $\Delta r=r_0$ and $r=r_0 n^{1/(D-2)}$.
Using $\rho(r)$ of Eq.\ (\ref{rho}), one has
\begin{equation}
m(N)=\frac{M}{\calN_0}\sum_{n=1}^N n\cdot\exp\left[-\alpha n^{2/(D-2)}\right]~,
\label{mN2}
\end{equation}
where 
\begin{equation}
\calN_0\equiv\sum_{n=1}^\infty n\cdot\exp\left[-\alpha n^{2/(D-2)}\right]~.
\end{equation}
Now the metric function for the discrete case is
\begin{equation}
h(N)
=1-\left(\frac{1}{\sqrt{\pi}M_* r_0}\right)^{D-3}
 \frac{8\Gamma\left(\frac{D-1}{2}\right)}{D-2}\frac{M}{M_*}
 \frac{1} {\calN_0 N^{\frac{D-3}{D-2}}}
 \sum_{n=1}^N n\exp\left[-\alpha n^{2/(D-2)}\right]~.
\label{hN}
\end{equation}
To find the minimum horizon radius $r_0$, it is required that 
 $h(N=1)=h'(N=1)=0$.
Since obtaining the analytic form of $h'(N)$ is very hard, we require that
$h(N=1)=0$ be the global minimum.
Numerical results for $r_0$ and the minimum value of $M$, $M_{min}$,and $m(N=1)$ is summarized
in Table \ref{T1}.
In this analysis we fix $M_*=1/\sqrt{\theta}=1$ TeV.
\begin{table}
\begin{tabular}{c|ccccccc}\hline
$D$ & $5$ & $6$ & $7$ & $8$ & $9$ & $10$ & $11$ \\\hline
$r_0$ & $2.89$ & $3.22$ & $3.53$ & $3.80$ & $4.06$ & $4.30$ & $4.53$ \\
$M_{min}$ & $~~22.4~~$ & $~~2.20\times 10^2~~$ & $~~2.01\times 10^3~~$ & $~~1.75\times 10^4~~$ & 
   $~~1.47\times 10^5~~$ & $~~1.21\times 10^6~~$ & $~~9.73\times 10^6$ \\
$m(N=1)$ & $9.83$ & $70.2$ & $4.76\times 10^2$ & $3.14\times 10^3$ & $2.02\times 10^4$ & $1.28\times 10^5$ & 
   $~~8.07\times 10^5$ \\
$\calN_0$ & $0.282$ & $0.233$ & $0.188$ & $0.150$ & $ 0.118$ & $0.0923$ & $0.0716$   
   \\\hline
\end{tabular}
\caption{Values of $r_0$, $M_{min}$, $m(N=1)$ in units of $\sqrt{\theta}$, $M_*$, and $M_*$
respectively, and $\calN_0$.}
\label{T1}
\end{table}
One can also find the values of $\calN_0$ approximately by investigating the 
converging values for sufficiently large summation number.
The result is also shown in Table \ref{T1}.
Note that the effective black hole mass is $m(N)$, which is in general smaller than $M_N$.
In $D=5$ dimensions $M_{min}=22.4$ (TeV) is quite out of reach of the LHC energy, 
while $m(N=1)=9.83$ (TeV) is much smaller than $M_{min}$, but it is still beyond the current LHC energy.
This might be the reason why the LHC has not seen any clues of mini black holes so far \cite{CMS,ATLAS}.
Even the full energy (14TeV) running could cover only $m(N=1)$ of 5 dimensions.
It was pointed out in \cite{Gingrich} that for some other values of $M_*$ for continuous case, 
the LHC can directly probe the NC black hole regime.
But the present analysis shows that the failure of current black hole searches at the LHC does not mean that 
higher dimensional black holes are not possible.
Much more comprehensive studies on the issue would appear elsewhere.
\par
The fact that much heavier mass is required to form a NC black hole is a very stringent point compared to the ordinary black holes.
The reason is that in the small length region $r\ll\sqrt{\theta}$  there exits radial pressure coming from the vacuum fluctuation \cite{Banerjee082}.
It acts against the inward gravitational collapse of matter to prevent the curvature singularity at the origin.
Another point that should be noticed is that $r_0$ gets larger for higher dimensions.
It is closely related to the qunatization of Eq.\ (\ref{rn}).
Roughly speaking, one needs larger value of $r_0$ for larger $D$ to make the same order of $r$.
This feature is quite contrary to the continuous case \cite{Nicolini11}.
\par
\begin{figure}
\includegraphics{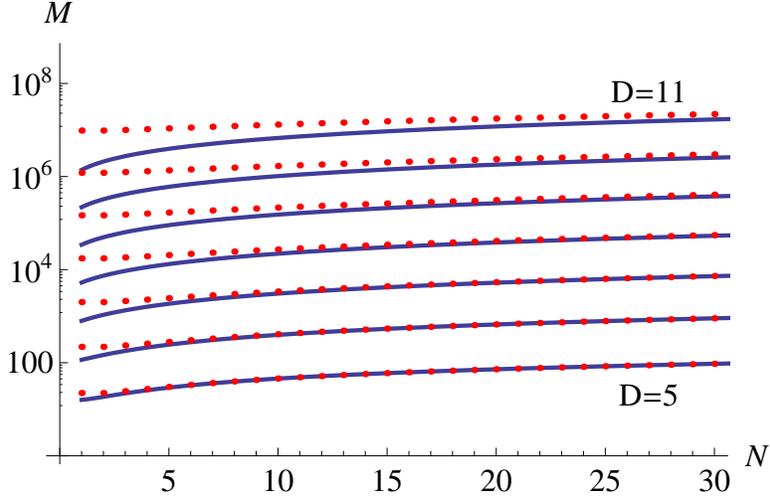}
\caption{Mass distributions (red dots) for different spacetime dimensions in units of $M_*$
($D=5,6,\cdots,11,$ from bottom to top).
Solid lines represent continuous mass functions.}
\label{massdist}
\end{figure}

Now the mass parameter $M$ is also quantized by the equation $h(N)=0$ with given
values of $r_0$ and $\calN_0$.
Explicitly,
\begin{equation}
 \frac{M}{M_*}=\frac{(D-2)\calN_0 (\sqrt{\pi}M_*r_0)^{D-3}N^{(D-3)/(D-2)}}{8\Gamma(\frac{D-1}{2})
 \sum^N n\cdot\exp\left[-\alpha n^{2/(D-2)}\right]}~,
\label{MN}
\end{equation}
from which one arrives at
\begin{equation}
 \frac{M_*}{M_N}N^\frac{D-3}{D-2}-\frac{M_*}{M_{N-1}}(N-1)^\frac{D-3}{D-2}
=\left(\frac{1}{\sqrt{\pi}M_* r_0}\right)^{D-3}\frac{8\Gamma\left(\frac{D-1}{2}\right)}{D-2}\frac{N}{\calN_0}
\exp\left(-\alpha\cdot N^\frac{2}{D-2}\right)~.
\label{Mrec}
\end{equation}
The mass spectrum $M_N$ is plotted in Fig.\ \ref{massdist}.
\par
In this figure, dotted plots are obtained from Eq.\ (\ref{MN}) for different $D$,
while solid lines are drawn from $h_{cont}(r)$.
In both dotted and solid plots, $r_0$'s are fixed by the requirement that $h(N=1)=0$ is the global minimum.
Note that two kinds of plots agree with each other very well for sufficiently large $N$,
which implies that the numerical results for the discrete mass spectrum are quite reliable.
However at low $N$, the mass spectra are very different (especially in higher dimensions) from the continuous ones.
In general in this region masses in discrete spectra are much heavier than those in continuous ones.
This is because for discrete spectra not all the masses are added up continuously, for example, in Eq.\ (\ref{mN1}).
Thus the discrete masses should be heavier to compensate for the sparse summation to form a black hole.
But for higher $N$ the difference between the two spectra becomes negligible 
since even in the discrete spectra many levels of masses are already added.
In this sense, heavier masses at low $N$ are good test bed to distinguish discrete spectra from continuous ones.
\par
The quantization rule for the effective mass $m(N)$ is rather simple.
According to Eq.\ (\ref{mN2}), 
\begin{equation}
 \frac{m(N)}{M_*}=\left[\frac{(D-2) (\sqrt{\pi}M_*r_0)^{D-3}}{8\Gamma(\frac{D-1}{2})}\right]N^{(D-3)/(D-2)}~.
\label{mN3}
\end{equation}
This result is consistent with \cite{Dvali} where the black hole mass $M_{BH}$ is 
$M_{BH}/M_*\sim N^{(D-3)/(D-2)}$.
\par
In conclusion, we provided with the quantization rules for $D$-dimensional NC black holes based on the holographic principle.
NC geometry requires that there must be a minimum mass (and minimum horizon radius) to form a black hole,
and it is assumed that spacetime is quantized such that $D$-dimensional surface is divided by fundamental area defined by the
minimum horizon radius.
Consequently the discrete mass spectra are obtained and they show different features from continuous ones at lower levels
while they are consistent with other results for higher levels.
The minimum mass scale for NC black hole formation is rather high and gets higher for discrete spectra, 
and this might be the reason why the LHC could not see any clue of mini black holes up to now.
\begin{acknowledgments}
The author thanks Yeong Gyun Kim for his hospitality and helpful discussions during the author's staying in GNUE
where this work was finalized.
This work was supported by NRF grant funded by MEST(No. 2011-0029758).
\end{acknowledgments}

\end{document}